\documentclass[12pt]{article}

\def\Title#1{\begin{center} {\Large #1 } \end{center}}
\def\Author#1{\begin{center}{ \sc #1} \end{center}}
\def\Address#1{\begin{center}{ \it #1} \end{center}}

\newenvironment{Abstract}{\begin{quotation}  }{\end{quotation}}
\newenvironment{Presented}{\begin{quotation} \begin{center} 
			PRESENTED AT\end{center}\bigskip 
		\begin{center}\begin{large}}{\end{large}\end{center} \end{quotation}}
\usepackage{graphicx}
\usepackage{bm}
\usepackage{txfonts}
\usepackage{hyperref}
\usepackage{epstopdf}
\usepackage{slashed}
\usepackage{rotating}
\usepackage{float}
\usepackage{array}
\date{\today}
\newcommand{\be}{\begin{eqnarray}}
	\newcommand{\ee}{\end{eqnarray}}

\newcommand{\bfp}{{\bf p}_{\perp}}

\newcommand{\Dp}{{\bf \Delta}_{\perp}}

\usepackage{xcolor}
\begin{document}
\begin{titlepage}
\vfill
\Title{Analysis of the higher twist GTMD $F_{31}$ for proton in the light-front quark-diquark model}
\vfill
\Author{ Shubham Sharma\textsuperscript{1},  Harleen Dahiya\textsuperscript{1}}
\Address{\bf $^1$ Department of Physics, Dr. B.R. Ambedkar National
	Institute of Technology, Jalandhar, 144027, India}
\vfill
\begin{Abstract}
In the light-front quark-diquark model (LFQDM), the higher twist generalized transverse momentum dependent distribution (GTMD) $F_{31}(x, {\bf p_\perp},{\bf \Delta_\perp})$ for the proton has been analyzed. We have derived the GTMD overlap equation by the analysis of GTMD correlator, employing the light-front wave functions in both the scalar and vector diquark situations. With the relevant 2-D and 3-D figures, the behavior of GTMD $F_{31}(x, {\bf p_\perp},{\bf \Delta_\perp})$ with variations in its variables has been illustrated. Further, on applying the transverse momentum dependent distribution (TMD) limit on GTMD $F_{31}(x, {\bf p_\perp},{\bf \Delta_\perp})$, the expression of TMD $f_3(x, {\bf p_\perp})$ has been obtained.
\end{Abstract}
\vfill
\begin{Presented}
DIS2023: XXX International Workshop on Deep-Inelastic Scattering and
Related Subjects, \\
Michigan State University, USA, 27-31 March 2023 \\
     \includegraphics[width=9cm]{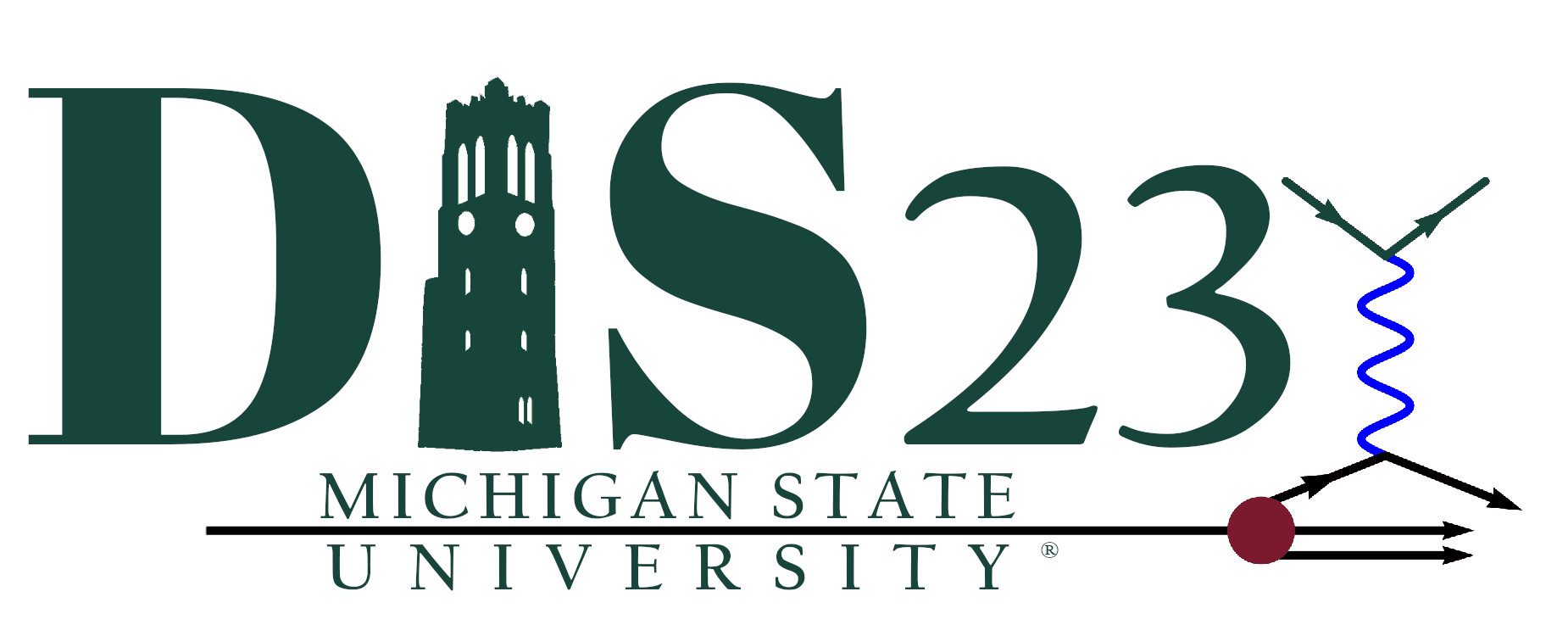}
\end{Presented}
\vfill
\end{titlepage}
\section{Introduction}
Understanding the surroundings have always been the nectar to the thirst of human consciousness. Quantum chromodynamics (QCD) is the branch of science concerning with the knowledge of the smallest building blocks of the universe which are partons. To describe the structure of hadron in terms of partons distributions like transverse momentum dependent parton distributions (TMDs) and generalized parton distributions (GPDs) came into picture. These distribution functions are related to the Drell-Yan (DY), deeply virtual Compton scattering (DVCS) and semi-inclusive deep inelastic scattering (SIDIS) processes. The enormous quantity of information which can be obtained from the correlator depicting the high-energy processes has been encrypted in generalized transverse momentum-dependent distributions (GTMDs). Functions at different twist have been defined depending on the helicity of hadron in the initial and the final state of the process. Lately, higher twist distributions have been the concerning topic of interest \cite{sstwist4,sstwist3}. In this work we have discussed about the higher twist GTMD $F_{31}$ for proton in the light-front quark-diquark model (LFQDM).
\section{Light-Front Quark-Diquark Model (LFQDM) \label{secmodel}}
In the LFQDM \cite{Maji:2016yqo}, proton is assumed to be the combination of active quark and a diquark observer. The spin-flavor $SU(4)$ structure is possessed by a proton and its description includes the amalgamation of isoscalar-scalar diquark singlet $|u~ S^0\rangle$, isoscalar-vector diquark $|u~ A^0\rangle$ and isovector-vector diquark $|d~ A^1\rangle$ states \cite{Maji:2016yqo}. 
The\\ smacked quark's ($p$) and diquark ($P_X$) momentum can be expressed as $p \equiv \big(xP^+, p^-,\bfp \big)$ and
$P_X \equiv \big((1-x)P^+,P^-_X,-\bfp\big)$ respectively. The expansion of Fock-state in the two particle case for $J^z =\pm1/2$ for the scalar $|\nu~ S\rangle^\pm $ and vector diquark $|\nu~ A \rangle^\pm$ in the situation of two particles can be expressed as \cite{Maji:2016yqo}
\begin{eqnarray}
	|\nu~ S\rangle^\pm & =& \int \frac{dx~ d^2\bfp}{2(2\pi)^3\sqrt{x(1-x)}} \Bigg[ \psi^{\pm(\nu)}_{+}(x,\bfp)\bigg|+\frac{1}{2},~s; xP^+,\bfp\bigg\rangle \nonumber \\
	&+& \psi^{\pm(\nu)}_{-}(x,\bfp) \bigg|-\frac{1}{2},~s; xP^+,\bfp\bigg\rangle\Bigg],\label{fockSD}\\
	|\nu~ A \rangle^\pm & =& \int \frac{dx~ d^2\bfp}{2(2\pi)^3\sqrt{x(1-x)}} \Bigg[ \psi^{\pm(\nu)}_{++}(x,\bfp)\bigg|+\frac{1}{2},~+1; xP^+,\bfp\bigg\rangle \nonumber\\
	&+& \psi^{\pm(\nu)}_{-+}(x,\bfp)\bigg|-\frac{1}{2},~+1; xP^+,\bfp\bigg\rangle +\psi^{\pm(\nu)}_{+0}(x,\bfp)\bigg|+\frac{1}{2},~0; xP^+,\bfp\bigg\rangle \nonumber \\
	&+& \psi^{\pm(\nu)}_{-0}(x,\bfp)\bigg|-\frac{1}{2},~0; xP^+,\bfp\bigg\rangle + \psi^{\pm(\nu)}_{+-}(x,\bfp)\bigg|+\frac{1}{2},~-1; xP^+,\bfp\bigg\rangle \nonumber\\
	&+& \psi^{\pm(\nu)}_{--}(x,\bfp)\bigg|-\frac{1}{2},~-1; xP^+,\bfp\bigg\rangle  \Bigg].\label{fockVD}
\end{eqnarray}
Here, the flavor index $\nu ~=u$ (for the case of scalar) and   $\nu ~=u,d$ (for the case of vector)). We have  $|\lambda_q,~\lambda_{Sp};  xP^+,\bfp\rangle$ representing the two particle state with quark and spectator diquark helicity of $\lambda_q=\pm\frac{1}{2}$ and $\lambda_{Sp}$ respectively. The helicity of spectator for scalar diquark is $\lambda_{Sp}=\lambda_{S}=0$ (singlet) and that for vector diquark is $\lambda_{Sp}=\lambda_{D}=\pm 1,0$ (triplet).
With the possibility of the diquarks being a scalar ($\rm{S}$) or a vector ($\rm{A}$), the LFWFs have been listed when $J^z=\pm1/2$ in Table \ref{tab_LFWF}.
\begin{table}[h]
	\centering 
	\begin{tabular}{ |p{0.6cm}|p{1.1cm}|p{1.0cm}|p{0.45cm} p{4.3cm}|p{0.45cm} p{4.3cm}|  }
		\hline
		&~~$\lambda_q$~~&~~$\lambda_{Sp}$~~&\multicolumn{2}{c|}{LFWFs for $J^z=+1/2$} & \multicolumn{2}{c|}{LFWFs for $J^z=-1/2$}\\
		\hline
		~$\rm{S}$&~~$+1/2$~~&~~$~~0$~~&~~$\psi^{+(\nu)}_{+}$~&~~$=~N_S~ \varphi^{(\nu)}_{1}$~~&~~$\psi^{-(\nu)}_{+}$~&~~$=~N_S \bigg(\frac{p^1-ip^2}{xM}\bigg)~ \varphi^{(\nu)}_{2}$~~  \\
		&~~$-1/2$~~&~~$~~0$~~&~~$\psi^{+(\nu)}_{-}$~&~~$=~-N_S\bigg(\frac{p^1+ip^2}{xM} \bigg)~ \varphi^{(\nu)}_{2}$~~&~~$\psi^{-(\nu)}_{-}$~&~~$=~N_S~ \varphi^{(\nu)}_{1}$~~   \\
		\hline
		&~~$+1/2$~~&~~$+1$~~&~~$\psi^{+(\nu)}_{+~+}$~&~~$=~~N^{(\nu)}_1 \sqrt{\frac{2}{3}} \bigg(\frac{p^1-ip^2}{xM}\bigg)~  \varphi^{(\nu)}_{2}$~~&~~$\psi^{-(\nu)}_{+~+}$~&~~$=~~0$~~  \\
		~~~~&~~$-1/2$~~&~~$+1$~~&~~$\psi^{+(\nu)}_{-~+}$~&~~$=~~N^{(\nu)}_1 \sqrt{\frac{2}{3}}~ \varphi^{(\nu)}_{1}$~~&~~$\psi^{-(\nu)}_{-~+}$~&~~$=~~0$~~   \\
		~$\rm{A}$&~~$+1/2$~~&~~$~~0$~~&~~$\psi^{+(\nu)}_{+~0}$~&~~$=~~-N^{(\nu)}_0 \sqrt{\frac{1}{3}}~  \varphi^{(\nu)}_{1}$~~&~~$\psi^{-(\nu)}_{+~0}$~&~~$=~~N^{(\nu)}_0 \sqrt{\frac{1}{3}} \bigg( \frac{p^1-ip^2}{xM} \bigg)~  \varphi^{(\nu)}_{2}$~~   \\
		&~~$-1/2$~~&~~$~~0$~~&~~$\psi^{+(\nu)}_{-~0}$~&~~$=~~N^{(\nu)}_0 \sqrt{\frac{1}{3}} \bigg(\frac{p^1+ip^2}{xM} \bigg)~ \varphi^{(\nu)}_{2}$~~&~~$\psi^{-(\nu)}_{-~0}$~&~~$=~~N^{(\nu)}_0\sqrt{\frac{1}{3}}~  \varphi^{(\nu)}_{1}$~~   \\
		&~~$+1/2$~~&~~$-1$~~&~~$\psi^{+(\nu)}_{+~-}$~&~~$=~~0$~~&~~$\psi^{-(\nu)}_{+~-}$~&~~$=~~- N^{(\nu)}_1 \sqrt{\frac{2}{3}}~  \varphi^{(\nu)}_{1}$~~   \\
		&~~$-1/2$~~&~~$-1$~~&~~$\psi^{+(\nu)}_{-~-}$~&~~$=~~0$~~&~~$\psi^{-(\nu)}_{-~-}$~&~~$=~~N^{(\nu)}_1 \sqrt{\frac{2}{3}} \bigg(\frac{p^1+ip^2}{xM}\bigg)~  \varphi^{(\nu)}_{2}$~~   \\
		\hline
	\end{tabular}
	\caption{The LFWFs for both diquark cases when $J^z=\pm1/2$, for various values of helicities of smacked quark $\lambda_q$ and the spectator diquark $\lambda_{Sp}$. The normalization constants are $N_S$, $N^{(\nu)}_0$ and $N^{(\nu)}_1$.}
	\label{tab_LFWF} 
\end{table}
The general form of LFWFs $\varphi^{(\nu)}_{i}=\varphi^{(\nu)}_{i}(x,\bfp)$ in the table has been obtained from the prediction of soft-wall AdS/QCD, and the establishment of parameters $a^\nu_i,~b^\nu_i$ and $\delta^\nu$ have been followed from Ref. \cite{Maji:2016yqo}. We have
\begin{eqnarray}
	\varphi_i^{(\nu)}(x,\bfp)=\frac{4\pi}{\kappa}\sqrt{\frac{\log(1/x)}{1-x}}x^{a_i^\nu}(1-x)^{b_i^\nu}\exp\Bigg[-\delta^\nu\frac{\bfp^2}{2\kappa^2}\frac{\log(1/x)}{(1-x)^2}\bigg].
	\label{LFWF_phi}
\end{eqnarray}
The model parameters and other constants have been given in Ref.  \cite{Maji:2016yqo}.

\section{Quark Correlator and Parameterization}
The unintegrated quark-quark GTMD correlator for proton at the fixed light-cone time $ z^+=0$, defined following Ref. \cite{Meissner:2009ww}, is given as
\begin{eqnarray} 
	W^{\nu [\Gamma]}_{[\Lambda^{N_i}\Lambda^{N_f}]}=\frac{1}{2}\int \frac{dz^-}{(2\pi)} \frac{d^2z_T}{(2\pi)^2} e^{ip.z} 
	\langle P^{f}; \Lambda^{N_f} |\bar{\psi}^\nu (-z/2)\Gamma \mathcal{W}_{[-z/2,z/2]} \psi^\nu (z/2) |P^{i};\Lambda^{N_i}\rangle \bigg|_{z^+=0}\,,
	\label{corr}
\end{eqnarray} 
where $|P^{i};\Lambda^{N_i}\rangle $ and $|P^{f}; \Lambda^{N_f}\rangle$ are the initial and final states of the proton respectively. and $\psi^\nu (\bar{\psi}^\nu)$ is the quark field operator. For the case of zero skewness i.e., $\xi=- \Delta^+/2P^+=0$, the GTMD correlator depends on the set of variables $(x,\bfp,\Dp,\theta)$, where $\theta$ being the angle between $\bfp$ and $\Dp$ plane. For convenience, the Wilson line, $\mathcal{W}_{[-z/2,z/2]}$ have considered it to be $1$. 
In the symmetric frame, proton's average momentum is stated as $P= \frac{1}{2} (P^{f}+P^{i})\equiv \big(P^+,\frac{M^2+\Dp^2/4}{P^+},\textbf{0}_\perp\big)$, while momentum transfer is represented as $\Delta=(P^{f}-P^{i})\equiv \big(0, 0,\Dp \big)$, where $P^{i}$ and $P^{f}$ are used to denote the initial and final four momenta of the proton respectively.
\par There are 4 quark GTMDs corresponding to the Dirac matrix structure $\Gamma=\gamma^-$ and it can be projected using Dirac spinor ${u}(P, \Lambda)$ as \cite{Meissner:2009ww}
\begin{equation}
	W_{[\Lambda^{N_i}\Lambda^{N_f}]}^{[\gamma^-]}
= \frac{M}{2(P^+)^2} \, \bar{u}(P^{f}, \Lambda^{N_F}) \, \bigg[
	F_{3,1}
	+ \frac{i\sigma^{i+} p_T^i}{P^+} \, F_{3,2}
	+ \frac{i\sigma^{i+} \Delta_T^i}{P^+} \, F_{3,3}   + \frac{i\sigma^{ij} p_T^i \Delta_T^j}{M^2} \, F_{3,4}
	\bigg] u(P^{i}, \Lambda^{N_i})
	\,, \label{par} \end{equation}
The indices $i$ and $j$ are used to denote transverse directions.

\section{Result and Discussion}\label{secresults}
By solving Eq. \ref{corr} and Eq. \ref{par} for GTMD $F_{3,1}$ via the use of LFWFs in Table \ref{tab_LFWF}, we get
%
\begin{eqnarray} 
	%
	%
	F_{3,1}^{\nu} &=&  \frac{1}{16 \pi^3 x^2 M^2}\Bigg[{{C_{S}^{2} N_s^2}+{C_{A}^{2}}\bigg(\frac{1}{3} |N_0^\nu|^2+\frac{2}{3}|N_1^\nu|^2 \bigg)}\Bigg] \Bigg[ \bigg(m^2+\bfp^2-\frac{\Dp^2}{4} \bigg) \varphi_1^{(\nu)}\varphi_1^{(\nu)}\nonumber\\
	&&+ \bigg(\Big(m^2+\bfp^2-\frac{\Dp^2}{4} \bigg)\bigg(\bfp^2-(1-x)^2\frac{\Dp^2}{4} \bigg)+(1-x)\bigg(\bfp^2 \Dp^2- (\bfp \cdot \Dp)^2 \Big)\bigg) \frac{\varphi_2^{(\nu)}\varphi_2^{(\nu)}}{x^2 M^2}\nonumber\\
	&&+ m (1-x)\Dp^2 \frac{\varphi_1^{(\nu)}\varphi_2^{(\nu)}}{x M}\Bigg]. \label{f31s}
\end{eqnarray}
\begin{figure*}
	\centering
	\begin{minipage}[c]{0.98\textwidth}
		(a)\includegraphics[width=7.0cm]{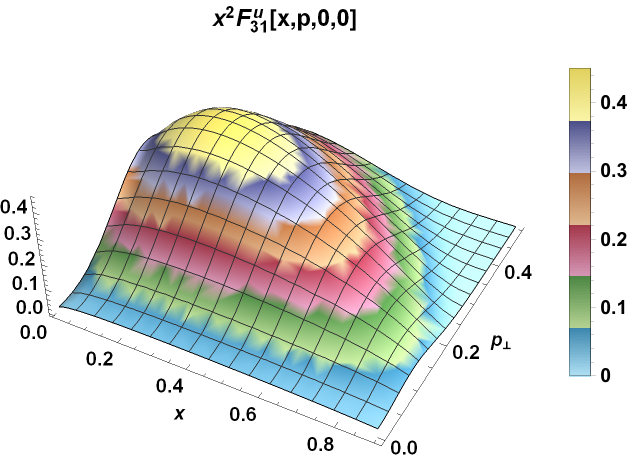}
		\hspace{0.05cm}
		(b)\includegraphics[width=7.0cm]{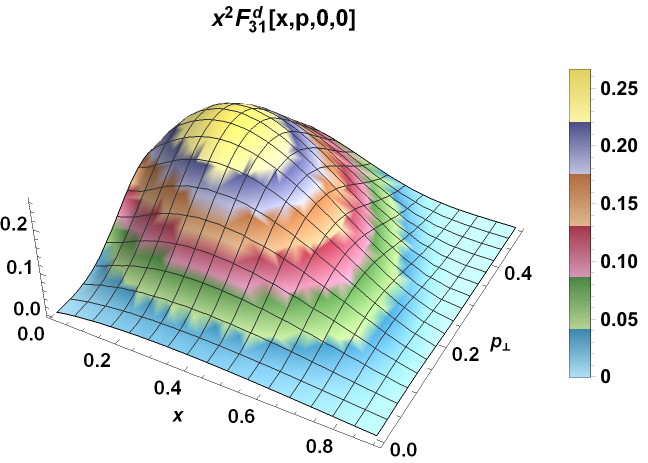}
		\hspace{0.05cm}
	\end{minipage}
	\caption{\label{fig3d} The twist-4 GTMD		
		$x^2 F_{3,1}^{\nu}(x, {\bf p_\perp},{\bf \Delta_\perp},\theta)$
		plotted with respect to $x$ and ${\bfp}$ for $\Dp= 0~\mathrm{GeV}$. The left and right column correspond to $u$ and $d$ quarks sequentially.}
\end{figure*}
\begin{figure*}
	\centering
	\begin{minipage}[c]{0.98\textwidth}
		(a)\includegraphics[width=7.0cm]{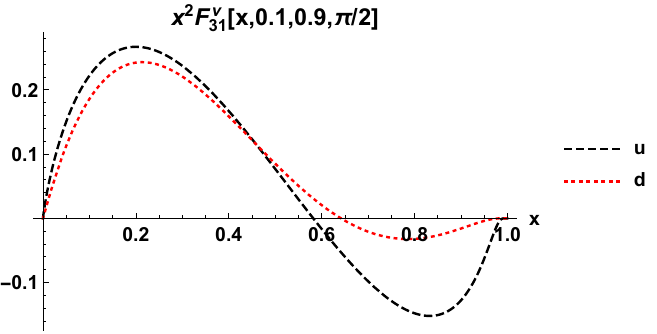}
		\hspace{0.05cm}
		(b)\includegraphics[width=7.0cm]{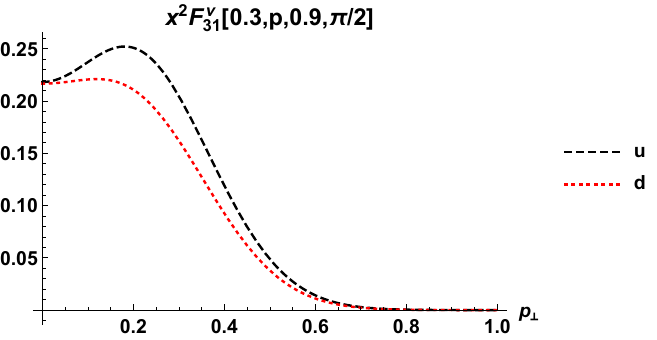}
		\hspace{0.05cm}
		(c)\includegraphics[width=7.0cm]{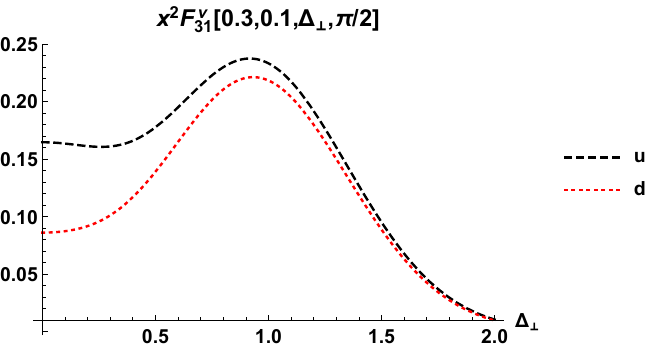}
		\hspace{0.05cm}
		(d)\includegraphics[width=7.0cm]{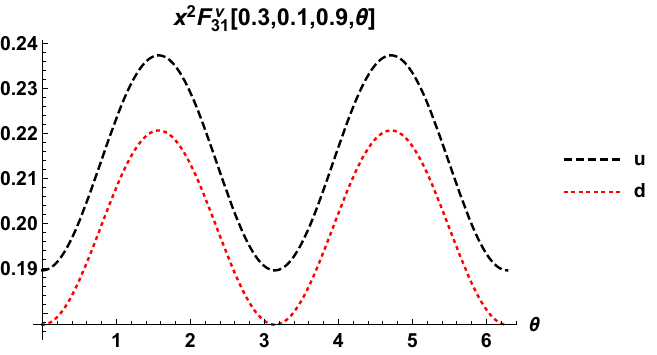}
		\hspace{0.05cm}
	\end{minipage}
	\caption{\label{fig2d} The twist-4 GTMD 		
		$x^2 F_{3,1}^{\nu}(x, {\bf p_\perp},{\bf \Delta_\perp},\theta)$
		variation individually with respect to $x$, ${\bfp}$, $\Dp$ and $\theta$.}
\end{figure*}
At the TMD limit $\Delta=0$ and the twist-4 T-even TMD can be obtained from GTMD as \cite{Meissner:2009ww}
\begin{eqnarray}
	F_{3,1}^{\nu}(x, {\bf p_\perp},0,\theta)& = &f_3^{\nu}(x,\bfp) \, \label{tmd1}  
\end{eqnarray}
In Fig. \ref{fig3d}, the GTMD $x^2 F_{3,1}^{\nu}(x, {\bf p_\perp},{\bf \Delta_\perp},\theta)$ has been plotted for both $u$ and $d$ quarks. To study this multidimensional function, the variation of GTMD $x^2 F_{3,1}^{\nu}(x, {\bf p_\perp},{\bf \Delta_\perp},\theta)$ for both quarks with its variables individually has been shown in Fig. \ref{fig2d}. The maximum possibility to obtain this kind of distribution exists when the quark carry at $20\%$ and $80\%$ of the longitudinal momentum from its parent. In Fig. \ref{fig2d} (b) and \ref{fig2d} (c), the trend at higher values appears logical as it is difficult to obtain quark carrying very high momentum and also it is less possible to observe very high momentum transfer between the initial and final states. From Eq. \ref{f31s}, it has been observed that when the ${\bfp} \parallel {\Dp}$, zero contribution from this term $\big(\bfp^2 \Dp^2- (\bfp \cdot \Dp)^2 \big)$ is obtained which leads to the minimum value of GTMD as evident from Fig. \ref{fig2d} (d). In all figures the amplitude of $u$ quarks is always greater than that of $d$ quarks.
 \section{Conclusion}\label{seccon}
 We have solved the quark-quark GTMD correlator at twist-4 for the calculation of GTMD $F_{3,1}^{\nu}$. We have observed that the presence of non-zero momentum transfer between the initial and the final states is the sole reason behind the comparatively lower possibility of GTMD over TMD $f_3^{\nu}$. Although the process contributing to GTMDs is not known yet but being mother distributions, it play significant role in the understanding of DY, SIDIS and DVCS processes by providing GPDs and TMDs.
 \par To understand the behavior of collision process in the future electron-ion collider when proton is in different polarized states, knowledge of distributions at higher twist is required. In future, it would be amazing to include the gluon and sea quark contributions in distributions to establish some model independent relations.
 
 \section{Acknowledgement}
 H.D. would like to thank the Science and Engineering Research Board, Department of Science and Technology, Government of India through the grant (Ref No.TAR/2021/000157) under TARE scheme for financial support.

\end{document}